\documentclass{article}
\begin{document}
\title{A note on the heat kernel coefficients for nonminimal operators}
\author{B. Ananthanarayan\\
Centre for High Energy Physics,\\
Indian Institute of Science, Bangalore 560 012, India\\
e-mail: anant@cts.iisc.ernet.in}
\date{August 19, 2008}
\maketitle

\vskip 3cm

\begin{abstract}
We consider certain results for the heat kernel
of nonminimal operators.
The general expressions provided by Gusynin and Kornyak
resulting from symbolic computation programmes for
$n$ dimensions are evaluated for 4 dimensions which are 
checked against results given by Barvinsky and Vilkovisky. 
We also check that the results in flat space are consistent with earlier
results of Guendelmen et al. 
We then consider a powerful construction of 
the Green function of a nonminimal operator by Shore for covariantly
constantly gauge fields in flat spacetime, and employ dimensional
arguments to produce a check on the
gauge parameter dependence of a certain coefficient.  
The connection of the results
for heat kernel coefficients emanating from the  construction
of Shore, to those from other techniques is hereby established
for the first time.  
\end{abstract}

\medskip


\medskip


\newpage

\section{Introduction}

A traditional approach to the evaluation of the divergent part
of the generating functional of Green functions in field theory is
the well-known heat kernel method, for a recent review see
ref.~\cite{Vassilevich}.  The coefficients in the asymptotic
expansion of the `diagonal' heat kernel elements are the
well-known Seeley-DeWitt coefficients. These are typically obtained
for so-called `minimal' operators of the type $(-D^\mu D_\mu+X)$.  Non-minimal
operators typically involve bi-linears of the type $(-g_{\mu\nu}D^\rho D_\rho
-(1/\alpha-1)D_\mu D_\nu+X_{\mu\nu})$, 
where the minimal case is obtained with $\alpha=1$.

In quantum field theory, the divergent part of the one-loop
generating functional of Green functions may be expressed
in terms of the second Seeley-DeWitt coefficient of 
certain differential operators.  The subject is by now
standard and is discussed in standard textbooks.  Given
a Lagrangian field theory, there is associated with this
a differential operator denoted by ${\cal D}$ such that 
the one-loop generating function $\Gamma^{(1)}$ in
the neighbourhood of $n=4$ is given
in dimensional regularization by 
\begin{eqnarray}
& \displaystyle
{i\over 2}\log \,{\rm det} {\cal D}=-\int d^n x\, 
{1\over (n-4)} {\rm tr} E_4   &
\end{eqnarray} 
where  $E_4$ is the
second Seeley-DeWitt or heat kernel coefficient
of the differential operator ${\cal D}$.

Typical differential operators that are considered are
of the minimal kind.
Non-minimal operators arise in gauge field theories in covariant
gauges in general, where $\alpha$ is the gauge parameter.
For the reasons mentioned above, the parameter is set equal to
unity, which corresponds to the Feynman gauge\footnote{Note, for instance,
that the divergent part of the one-loop generating functional in 
chiral perturbation theory with
virtual pions has been computed only in the Feynman gauge~\cite{CHPT}.}.   
Differential operators also appear in curved spacetime
which involve the Ricci tensor, curvature tensor and the scalar
curvature.  
The traditional method of evaluating the corresponding heat kernel
coefficients going under the name of the method of DeWitt
does not work for the nonminimal case.  
Techniques used for such nonminimal operators 
go under the name of the method of Widom.
There are techniques advanced in the literature which
provide algorithms based on the method of Widom to compute the
heat kernel coefficients, refs.~\cite{Gusynin1,Gusynin2,Gusynin3}.
Note that nonminimal operators have also been
considered in the context of noncommutative field theories~\cite{Noncomm}.

Results have been presented by
Barvinsky and Vilkovisky (BV) in ref.~\cite{Barvinsky} with curvature, 
where some of the results crosscheck those that
were presented by in earlier literature.
More recently, Pronin and Stepanyantz (PS)
in ref.~\cite{Pronin1} have also considered the
nonminimal case and find results consistent with those in
ref.~\cite{Barvinsky}.  The heat kernel coefficient
corresponding to what is a surface term is not given by
PS.  Even more recently these operators have been
studied by Gusynin and Kornyak (GK), in ref.~\cite{comprehensive}
using symbolic computation and including the tensor
denoted by $W_{ij}$ to account for gauge fields,
and results have been provided for
general case of $n$ dimensions.  
However, no attempt has been provided to compare the
results from this to those of BV and
PS, although it is a straightforward exercise.
Here we provide such a comparison, as it is very
important to check the results in every possible manner.  
Furthermore, it is important
to look for an important crosscheck on the coefficient of
the bilinears involving the gauge field strength tensor, 
and in particular of its dependence on the parameter
$\alpha$.  Results have been
provided by Guendelman et al.~\cite{Guendelman}, which
are also based on symbolic computation methods.

In this regard, we show here that an elegant
analytical approach is also available to accomplish this
goal:  we look at a
completely different solution present in the literature
which has not attracted attention to the best
of our knowledge.  This is the general case considered by
Shore, ref.~\cite{Shore}, for the case of a covariant
gauge, but in flat spacetime.  It is shown here that a study of his
construction can provides a consistency check on the results
obtained from the general expressions of GK
for the case of four dimensions for the bilinear in gauge field
strength.  A different approach that also provides
a proof of scuh gauge parameter independence 
is given by Avramidi~\cite{Avramidi}.

In Sec.~\ref{GK} we present an evaluation of the heat kernel coefficients
for 4 dimensions from the general expressions given by
of GK.  We carefully compare the results given
by BV and PS.  It will turn
out that one of the coefficients in our evaluation remains untested
at this stage, but interestingly enough is independent of
$\alpha$ for $n=4$.  
This coefficient along another combination of
coefficients that is itself independent
of $\alpha$ for $n=4$, which we will discuss, 
will be related to the work of
Guendelman et al.,  who computed what is
effectively this combination using symbolic computation.  
Yet other coefficients are listed here for the first time for $n=4$.
We advance here an analytical argument
in Sec.~\ref{CS} where we consider in considerable detail
the construction of Shore and work out the consequences for the
heat kernel coefficients.  
In Sec.~\ref{Conc}
we provide a discussion on the results and
recapitulate the main results in this work.

\section{The Results of Gusynin and Kornyak for 4 dimensions}\label{GK}

A comprehensive treatment for the evaluation of the trace of the
second Seeley-DeWitt coefficient termed $E_4$ is
provided by GK~\cite{comprehensive}.
In this paper, 
the trace of the Seeley-DeWitt coefficient 
evaluated explicitly in curved background, and in arbitrary
gauge, and a list of  $C_i,\, i=1,...14$, is provided in
$n$ dimensions, in terms of a parameter $a$, where $a=1-1/\alpha$.
For all other definitions and conventions we refer to
the paper of GK.
Recall that the divergent part of the
generating function is given by 
the spacetime integral of

\newpage

\begin{eqnarray}\label{gkequation}
 & \displaystyle {\rm tr} E_4= (4 \pi)^{-n/2} \cdot
& \nonumber \\
& \displaystyle 
\left[
-C_1 D_i D^i X^j_j - C_2 D_i D_j X^{ij} - C_3 D_i D_j
X^{ji} + {C_4\over 2}(X^i_j X^j_j + X_{ij} X^{ij}) + 
 \right. & \\
& \displaystyle \left. 
C_5 X_{ij} X^{ji} + 
C_6 X_{ij} W^{ij} - C_7 W_{ij} X^{ij} + C_8 W_{ij}W^{ij}
+ C_9 R_{ijkl} R^{ijkl} - \right. & \nonumber \\
& \displaystyle \left. C_{10} R_{ij} X^{ij} - C_{11} R_{ij} R^{ij} +
C_{12} D_i D^i R + C_{13} R^2 - C_{14} R X^i_i \right] & \nonumber
\end{eqnarray}

\begin{center}
\begin{table}
\begin{tabular}{|c||c||c||}\hline\hline
Term & Coefficient & Value  \\ \hline\hline
$R_{ijkl} R^{ijkl} - 4 R_{ij} R^{ij} + R^2$ & $C_9$ & $-11/180$ \\ \hline
$R_{ij} R^{ij}$ & $(4 C_9 - C_{11})$ & $(5 \gamma^2+10 \gamma - 32)/120$
						         \\ \hline
$R^2$ & $(C_{13}-  C_9)$ 
& $(5\gamma^2+20 \gamma + 28)/240$          \\ \hline
$R_{ij} X^{ij}$ &  $-C_{10}$ & $-\gamma(\gamma+4)/12$         \\ \hline
$X_{ij} X^{ij}$ & $(C_4/2+ C_5)$ 
	& $(\gamma^2+6\gamma+12)/24$         \\ \hline
$R X^i_i$ & $-C_{14}$ & $-(\gamma^2+2\gamma+4)/24$         \\ \hline
$X^i_i X^j_j$ & $C_4/2$ & $\gamma^2/48$         \\ \hline
$W_{ij} W^{ij}$ & $C_8$ & $1/3$     \\  \hline \hline
\end{tabular}
\caption{List of irreducible basis of tensor bilinears appearing
in the divergent part of the one-loop generating functional, the
corresponding coefficients, and their values.}
\end{table}
\end{center} 

We introduce a further parameter
$\gamma\equiv a/(1-a)$ in order to have an effective
comparison with the results of BV.
We evaluate these for the case of 4 dimensions from the
general formulae of GK and tabluate 
the (combinations) of coefficients in Tables 1 and 2. 
The results expressed in Table 1 are grouped to effect an easy
comparison with known results in the literature.  In particular,
we are presenting those combinations of $C_4,\, C_5,\, 
C_9,\, C_{10},\, C_{11}$ and $C_{13}$ which appear in 
the work of BV.
The following may be noted:

\medskip

\noindent
(a) The sign convention for
$C_{10},\, C_{14}$ differs from that in BV.

\medskip

\noindent
(b) We regroup the terms to obtain the combination $(R_{ijkl} R^{ijkl}
-4 R_{ij} R^{ij} + R^2)$ (surface term).

\medskip

\noindent
(c) Our results are in
complete agreement with BV (also
with those of PS, while noting that the latter omit
the surface term).  Note that in BV the divergent part
of the generating functional involves $(\log \, L^2)$ (where
$L$ is a large momentum scale) and it may be noted that
one may map results obtained with cutoff regularization
with those in dimensional regularization by identifying
this with $-2/(n-4)$.

\medskip

\noindent
(d) The last entry in Table 1 is not present in BV
and needs to be verified independently, at least for
the dependence on the gauge parameter.

\medskip

\noindent
(e)
Despite the lack of details GK, one may try to compare
the results in GK with those of Guendelman et al.~\cite{Guendelman}.
In order to carry out a comparison with the results in the
work of Guendelman et al, 
the following may be readily noted: with the identification
$X_{ij}=-2 W_{ij}$, where $W_{ij}^{ab}=   f^{a b c} F^{c}_{ij}$
(see eq.~(2) in ref.~\cite{Guendelman}), the resulting coefficient
of $W_{ij} W^{ij}$ is given by
\begin{eqnarray} \label{guendelmaneq}
& \displaystyle (2 C_4 - 4 C_5 -2 C_6 +2 C_7 + C_8)=
{1\over 12} (-25 + n + \alpha^{n/2-2}) &
\end{eqnarray}
which is in agreement with eq.~(13) in ref.~\cite{Guendelman}.
Other terms in eq.~(\ref{gkequation}) for this case in 
flat space vanish due to reasons of symmetry. 
The checks with the results of BV provide a check on the
$\alpha$ independence of 
$(2 C_4 - 4 C_5 -2 C_6 +2 C_7)$, but that of $C_8$ can be
checked only from the above.
Thus, we show here for the first time the agreement of results
obtained by two independent groups, which constitutes an
important cross-check on the results.

\medskip

\noindent
(f)
Despite all the cross-checks carried out so far,
what is of interest to us here is to find an analytical
argument for the feature of gauge independence of the 
combination on the left hand side of eq.~(\ref{guendelmaneq}) for $n=4$. 
In order to facilitate this latter, we will turn to the construction of
Shore which is the subject of the next section.  

\bigskip

In Table 2 we present the values obtained for those coefficients
that do not appear in Table 1.  These have not, to the best of
our knowledge, appeared in the literature for 4 dimensions.\footnote{
Expressed differently, this is a $\log\, \alpha$ dependence of
some terms in the divergent part of the effective Lagrangian, which
has been noted in the context of 
resonance saturation in chiral perturbation theory~\cite{BM}.}   These
have a well-defined limit in the Feynman gauge ($\alpha=1,\, a=\gamma=0$):
$C_1=1/6,\, C_2=C_3=C_6=C_7=0,\, C_{12}=2/15$.
These do not appear in BV
as those accompanying $C_{1,2,3,12}$ vanish upon spacetime integration
and those accompanying $C_{6,7}$ do not appear when gauge fields
are not present.

\begin{small}
\begin{center}
\begin{table}
\begin{tabular}{|c||c||}\hline\hline
Coefficient & Value  \\ \hline\hline
$C_1$ &  $\{\gamma(-6+9 \gamma + 7 \gamma^2) + 6 (1-\gamma^2)\log(1+\gamma)\}/
						(36 \gamma^2)$  
\\ \hline
$C_2$ &  $\{\gamma(96+150 \gamma + 29 \gamma^2-6\gamma^3) - 
		6 (16+33\gamma+17\gamma^2)\log(1+\gamma)\}/
						(72 \gamma^2)$  
\\ \hline
$C_3$ &  $\{-\gamma(48+66  \gamma + 19 \gamma^2-6\gamma^3) + 
		6 (8 +15\gamma+ 7\gamma^2)\log(1+\gamma)\}/
						(72 \gamma^2)$  
\\ \hline
$C_6$ & $-\{\gamma(288+756 \gamma+654 \gamma^2+156 \gamma^3 - 27 \gamma^4 +
	4 \gamma^5)$  \\
& ~~~$-36 (1+\gamma)^2(8+9\gamma) \log(1+\gamma) \}/(288 \gamma^2(1+\gamma)$ 
				\\ \hline
$C_7$ & $-\{\gamma(288+756 \gamma+510 \gamma^2+12  \gamma^3 - 27 \gamma^4 +
	4 \gamma^5)$  \\
& ~~~$-36 (1+\gamma)^2(8+9\gamma) \log(1+\gamma) \}/(288 \gamma^2(1+\gamma)$ 
				\\ \hline
$C_{12}$ &  $\{\gamma(60+288 \gamma + 95 \gamma^2) - 
		30(2 +9\gamma+ 6\gamma^2)\log(1+\gamma)\}/
						(360\gamma^2)$  
\\ \hline
\end{tabular}
\caption{List of the remaining coefficients}
\end{table}
\end{center}
\end{small}

\section{Heat-Kernel coefficients from Shore's
construction}\label{CS}
Shore considers
the case of a covariantly constant field and 
obtains an explicit form for the entire heat kernel.
In terms of the heat kernel obtained for the minimal case, an
expression is provided for the nonminimal case as well.  
An expression for the one-loop divergence
as the logarithm of the determinant of the relevant operator.
For caveats regarding the use of the expressions in the
$R$ gauge and conventional Lorentz gauge which will not
affect our results, and for more details, we refer
to the paper of Shore.  A similar construction was also
considered earlier by Endo~\cite{Endo}.

There are several steps in the programme which is described below
in some detail, keeping in mind that precise definitions may
be found in the paper of Shore:  

\medskip

\noindent
(a)
The object of the study of Shore is the kernel
${\cal G}^{ab}_{\mu\nu}(x,y,;t,m^2)$ for the vector operator
${\cal D}^{ab}_{\mu\nu}$ for arbitrary $\alpha$.  It is
defined by
\begin{eqnarray}
& \displaystyle
(-D^2 g_{\mu\lambda} + (1-{1\over\alpha})D_\mu D_\lambda
 + 2 i g_R F_{\mu\lambda}
+ m^2 g_{\mu\lambda})^{ac} {\cal G}^{cb}_{\lambda\nu}(x,y,;t;m^2) & \nonumber \\
& \displaystyle
= -{\partial {\cal G}^{ab}_{\mu\nu}(x,y,;t;m^2)\over \partial t},
&
\end{eqnarray}
and satisfying the initial condition
\begin{eqnarray}
{\cal G}^{ab}_{\mu\nu}(x,y;0,m^2)=\delta_{ab}g_{\mu\nu}\delta(x,y)
\end{eqnarray}
The covariant derivative $D_\mu^{ab}\equiv \partial_\mu \delta^{ab} - i g
A_\mu^c t^c_{ab}$, where $A_\mu^c$ is the gauge field and $t^c_{ab}$
are the generators of the gauge group.

\medskip

\noindent
(b)
If the condition $D_\mu m^2=0$ is satisfied then
the kernel for nonzero mass factorizes into 
\begin{eqnarray}
{\cal G}_{\mu\nu}(x,y ;t;m^2)={\cal G}_{\mu\nu}(x,y;t;0) \exp(-m^2(y) t)
\end{eqnarray}

\medskip

\noindent
(c)
There is an ansatz that relates the solution for the nonminimal operator
to that of the minimal operator ($\alpha=1$) with $m^2=0$, for the
case of covariantly constant fields.  The heat kernel for the minimal
operator with $m^2=0$ is denoted by $\overline{{\cal G}}_{\mu\nu}(x,y;t)$,
and the corresponding Green's function is denoted by
$\overline{G}(x,y)$.  Armed with this, the function ${\cal H}_{\mu\nu}$
is constructed and the desired heat kernel for the zero mass case is
constructed via
\begin{eqnarray}
{\cal G}_{\mu\nu}(x,y;t,0)=\overline{{\cal G}}(x,y;t)+D_\mu D_\lambda   
\left\{{\cal H}_{\lambda\nu}(x,y;t)-{\cal H}_{\lambda\nu}(x,y;t/\alpha)
\right\}
.
\end{eqnarray}

\medskip

\noindent
(d)
This expression has an remarkable property in that the $\alpha$
dependence factors out completely.  
An explicit expression for the heat kernel of the minimal operator
with $m^2=0$ for the case of the covariantly constant gauge field
strength is provided, and eventually an expression for the logarithm
of the determinant of the operator.

\medskip

\noindent
(e)
Consider now the result presented in eq.(4.67) in ref.~\cite{Shore}.  
In this expression, for our purposes it suffices to suppress the
trace over the gauge indices (Tr), and instead introduce
a constant $C$, and inserting the spacetime trace
(tr) for the case of $n=4$, 
we write down the schematic expression for the logarithm of
the determinant of the differential operator as:
\begin{eqnarray}
& \displaystyle {1\over C}\log \, {\rm det}\,
 {\cal D}=- (4\pi)^{-n/2} \int d^n x \cdot & 
\nonumber \\
& \displaystyle
\int_0^\infty  dt\, t^{-1-n/2} \left[(g F t)^2/\sin^2(g F t) \left\{
(4 \cos(2 g F t) - 1) e^{- m^2 t} + e^{- \alpha m^2 t}\right\} \right]
&
\end{eqnarray}
The divergent part is now obtained by expanding out
the parts of the integrand that do not involve the exponentials
in powers of $t$.  Recalling that
\begin{eqnarray}
& \displaystyle
\int_0^\infty dt \, t^{r-1-n/2} e^{-m^2 t}= \Gamma(r-n/2) (m^2)^{n/2-r}, & \\
& \displaystyle
\Gamma(-k+\epsilon)= {(-1)^k\over k!}\left({1\over \epsilon}+ ...\right),
\, k=0,1,2,... & 
\end{eqnarray}
we can readily see that divergent part in the spacetime
integrand now reads for the case of four dimensions:
\begin{eqnarray}\label{shorehk}
& \displaystyle
-{1\over 2} {1\over 16 \pi^2} {1\over n-4} (3+\alpha^2) m^4 +
{1\over 3} {g^2\over 16 \pi^2} {1\over n-4} 20 F^2.
\end{eqnarray}
It may be recalled here that the residue of the pole at 4 in the spacetime
integrand of 
$(-1/2) \log {\rm det}\, {\cal D}$ is 
the trace of the 2nd Seeley-DeWitt coefficient, keeping in
mind that Shore employs the Euclidean generating functional. 

\medskip

\noindent (f)
What is of note above is that the $F^2$ piece is independent of $\alpha$
which may be inferred from dimensional considerations.  

\medskip

To summarize, what we obtain from the analysis of 
the construction of Shore is the
prediction that the $m^4$ piece in the divergent part is
proportional to $(3+\alpha^2)$,                                
and that the $F^2$ piece is independent of $\alpha$.

In order to make contact with the results of the previous section,
it may be readily checked that, up to the factor $C$,

\medskip

(A) for the case of $X_{ij}=m^2 g_{ij}$
we get back the $(3+\alpha^2)$ dependence for the coefficient for 
$m^4$ by evaluating $[16 (C_4/2+C_5)+2 C_4]$ from Table 1,

\medskip

(B) we find a simple justification for the $\alpha$ independence of 
the combination given in eq.~(\ref{guendelmaneq}). 

\newpage

\section{Discussion and summary}\label{Conc}
We have considered in some detail the implications of the results
given in the work of GK for four dimensions.  
The results check those of
BV, including the one result in the latter that
was not checked earlier by the results of PS,
namely that of the surface contribution.  In addition, we have 
considered the remarkable construction of Shore for the case of
covariantly constant fields, for which a complete construction
of the Green function for the nonminimal case is provided and
employ this to obtain the heat kernel coefficients for a simplified
representation.  The one
corresponding to the $m^4$ term is shown to have a $(3+\alpha^2)$
dependence which agrees with the results of GK
and that of BV and PS.
In addition the construction of Shore provides a simple dimensional
argument for why the $F^2$ term should be independent of $\alpha$.
This agrees with the observation of Guendelman et al. which was
found using symbolic computation.  
The remaining coefficients $C_{1,2,3,6,7,12}$ are
also evaluated in 4 dimensions. 

While it would be interesting
to demonstrate that the construction of Shore is indeed consistent
with the method of Widom in a formal manner, we have demonstrated
instead that the results from this construction are in agreement
with the heat kernel coefficients obtained from the Widom method.

\bigskip

\section{Acknowledgements:}  We thank the Department of Science and
Technology, Government of India for support during the course of
this work.  Special thanks are due to
H. Leutwyler for discussions.
Discussions with B. Moussallam, S. Nampuri, A. K. Nayak,
S. Mallik, J. Samuel, K. Shivaraj,
A. Upadhyay and S. Vaidya are acknowledged.
We thank Prof. M. Shaposhnikov at the Ecole Polytechnique F\'ed\'erale
de Lausanne for his hospitality when part of this work was done,
and the Indo-Swiss Bilateral Research Initiative for support.

\end{document}